\documentclass[aps,prl,twocolumn,superscriptaddress]{revtex4}
\usepackage{graphicx}
\usepackage{amssymb}
\usepackage{pifont}

\begin{document}



\title{Novel phase diagram for antiferromagnetism and superconductivity in pressure-induced heavy-fermion superconductor Ce$_2$RhIn$_8$ probed by In-NQR}

\author{M.~Yashima}
\affiliation{Department of Materials Engineering Science, Osaka University, Osaka 560-8531, Japan}
\affiliation{JST, TRIP (Transformative Research-Project on Iron Pnictides), Chiyoda, Tokyo 102-0075, Japan}
\author{S.~Taniguchi}
\affiliation{Department of Materials Engineering Science, Osaka University, Osaka 560-8531, Japan}
\author{H.~Miyazaki}
\affiliation{Department of Materials Engineering Science, Osaka University, Osaka 560-8531, Japan}
\author{H.~Mukuda}
\affiliation{Department of Materials Engineering Science, Osaka University, Osaka 560-8531, Japan}
\affiliation{JST, TRIP (Transformative Research-Project on Iron Pnictides), Chiyoda, Tokyo 102-0075, Japan}
\author{Y.~Kitaoka}
\affiliation{Department of Materials Engineering Science, Osaka University, Osaka 560-8531, Japan}
\author{H.~Shishido}
\altaffiliation{Present Address: Department of Physics, Kyoto University, Kyoto 606-8502, Japan}
\affiliation{Department of Physics, Graduate School of Science, Osaka University, Osaka 560-0043, Japan}
\author{R.~Settai}
\affiliation{Department of Physics, Graduate School of Science, Osaka University, Osaka 560-0043, Japan}
\author{Y.~\=Onuki}
\affiliation{Department of Physics, Graduate School of Science, Osaka University, Osaka 560-0043, Japan}
\affiliation{Advanced Science Research Center, Japan Atomic Energy Research Institute, Tokai, Ibaraki 319-1195, Japan}

\begin{abstract}
We present a novel phase diagram for the antiferromagnetism and superconductivity in Ce$_2$RhIn$_8$ probed by In-NQR studies under pressure ($P$). The quasi-2D character of antiferromagnetic spin fluctuations in the paramagnetic state at $P$ = 0 evolves into a 3D character because of the suppression of antiferromagnetic order for $P > P_{\rm QCP}\sim$ 1.36 GPa (QCP: antiferromagnetic quantum critical point). Nuclear-spin-lattice-relaxation rate $1/T_1$ measurements revealed that the superconducting order occurs in the $P$ range 1.36 -- 1.84 GPa, with maximum $T_c\sim$ 0.9 K around $P_{\rm QCP}\sim$ 1.36 GPa.
\end{abstract}
\vspace*{5mm}
\pacs{74.25.Ha; 74.62.Fj; 74.70.Tx; 75.30.Kz} 

\maketitle
\section{I. INTRODUCTION}

The heavy-fermion (HF) compounds CeIn$_3$ \cite{Mathur,Knebel} and CeTIn$_5$ (T = Co, Rh, Ir) \cite{Petrovic1,Hegger,Muramatsu,Petrovic2} revealed an intimate relationship between antiferromagnetism (AFM) and superconductivity (SC) \cite{Kitaoka}. CeIn$_3$ has a cubic crystal structure, and it is expected to exhibit the three-dimensional (3D) magnetic interaction. CeIn$_3$ is an antiferromagnet with $T_N=10$ K at ambient pressure ($P$ = 0), and AFM discontinuously collapses around $P_c$ = 2.46 GPa, suggesting that the quantum phase transition from AFM to paramagnetism (PM) is of the first order \cite{SKawasaki1,SKawasaki2}. SC appears in a narrow pressure range $P$ = 2.28 -- 2.65 around $P_c$, and $T_c$ reaches the maximum value ($\sim$ 0.25 K) at $P_c$. Non-Fermi liquid behaviors observed at pressures below $P_c$ evolve into Fermi-liquid behaviors at pressures that just exceed $P_c$. It was suggested that the first-order quantum phase transition is responsible for the occurrence of SC in CeIn$_3$ \cite{SKawasaki2}.

CeRhIn$_5$, which has a tetragonal crystal structure, is also an antiferromagnet with $T_N$ = 3.8 K at $P$=0 \cite{Hegger}. For CeRhIn$_5$, we have shown that the tetracritical point, where the AFM, AFM+SC, SC, and PM phases are in contact, exists at $P_{\rm tetra} \sim$ 1.98 GPa and $T_c$ reaches the maximum value ($\sim$ 2.2 K) at  approximately 2.5 GPa from the AFM quantum critical point (QCP), which lies at $P_{\rm QCP}\sim$ 2.1 GPa (see Fig. 5c) \cite{Yashima1}. In the region where $P$ exceeds 2.1 GPa, non-Fermi liquid behaviors, which were probed by the resistivity measurements \cite{Muramatsu}, were observed and NQR measurements revealed the development of AFM spin fluctuations \cite{Yashima1}. CeTIn$_5$, Ce$_2$TIn$_8$ and, CeIn$_3$ (T = Co, Rh, Ir) are a series of structurally related materials with chemical compositions of the form Ce$_m$TIn$_{3m+2}$ with $m$ = 1, 2, $\infty$, respectively. Ce$_2$TIn$_8$ enables us to study the relationship between the structure-based evolution of magnetic characteristics and the onset of unconventional SC in HF systems.

Ce$_2$RhIn$_8$ is an antiferromagnet with $T_N$ = 2.8 K at $P$ = 0 \cite{Bao}. The collinear antiferromagnetic structure with a magnetic wave vector $Q = (1/2, 1/2, 0)$ and a magnetic moment of 0.55 $\mu_B$ per Ce ion was reported from the neutron scattering measurements \cite{Bao}. The pressure-temperature ($P-T$) phase diagrams of Ce$_2$RhIn$_8$ reported thus far are based on resistivity, ac-susceptibility, and heat-capacity measurements \cite{Nicklas,Ohara,Ueda,Lengyel}. The resistivity measurements revealed that as $P$ increases, $T_N$ monotonously decreases down to 1.2 K at 1.5 GPa; further, SC occurs for $P >$ 1 GPa and exhibits the maximum $T_c$ ($T_c^{max} \sim$ 2 K) around 2.3 GPa. On the other hand, the heat-capacity measurements indicated that an AFM order survives up to $P$ = 1.65 GPa, but no anomalies that signal the onset of SC were observed. The previously reported NQR-$1/T_1$ measurement was performed to investigate the onset of SC with $T_c$ = 0.9 K at $P$ = 1.87 GPa \cite{Fukazawa}. In this context, a $P-T$ phase diagram for Ce$_2$RhIn$_8$ is not yet fully understood.

\begin{figure}[htbp]
\centering
\includegraphics[width=7cm]{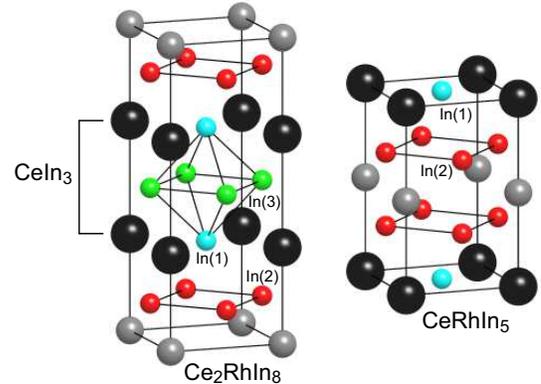}
\caption[]{\footnotesize  (Color online) (a) Crystal structures of Ce$_2$RhIn$_8$ and CeRhIn$_5$}
\label{crystal}
\end{figure}

\section{I\hspace{-.1em}I. Experimental procedure}

For obtaining NQR measurements, Ce$_2$RhIn$_8$ grown by the self-flux method was moderately crushed into a coarse powder to allow RF pulses to easily penetrate the sample. Hydrostatic pressure was applied using a NiCrAl-BeCu piston-cylinder cell filled with a Si-based organic liquid as the pressure-transmitting medium \cite{Kirichenko}. To calibrate the pressure at low temperatures, the shift in the $T_c$ of Sn metal was monitored by using the resistivity measurements. Figure~\ref{crystal} illustrates the crystal structure of Ce$_2$RhIn$_8$, which consists of alternating layers of CeRhIn$_5$ and CeIn$_3$. There are three In sites per unit cell, denoted by In(1), In(2), and In(3). In(1) and In(2) are located in the CeRhIn$_5$ layer, shown in Fig.~\ref{crystal}, and In(3) is located in the CeIn$_3$ layer. The measurements for the $^{115}$In-NQR ($I=9/2$) spectrum were mainly performed at the 3$\nu_Q$ transition at In(2) in Ce$_2$RhIn$_8$. Here, $\nu_Q$ is defined by the NQR Hamiltonian, $\mathcal{H}_Q$ = $(h\nu_Q/6)[3{I_z}^2-I(I+1)+\eta({I_x}^2-{I_y}^2)]$, where $\eta$ is the asymmetry parameter of the electric field gradient. Using $\nu_Q$ = 16.41 MHz and $\eta$ = 0.43, the NQR frequency of the 3$\nu_Q$ transition is estimated as 47.4 MHz for In(2) at $P$ = 0.  When an internal magnetic field $H_{\rm int}$ is present at the In site during the onset of AFM, the NQR Hamiltonian is perturbed by the Zeeman interaction, which is given by $\mathcal{H}_{\rm AFM}=-\gamma \hbar\vec{I} \cdot \vec{H}_{\rm int} + \mathcal{H}_Q$. A broadening of the NQR spectrum due to $H_{\rm int}$ signals the onset of AFM.

\begin{figure}[htbp]
\centering
\includegraphics[width=6.5cm]{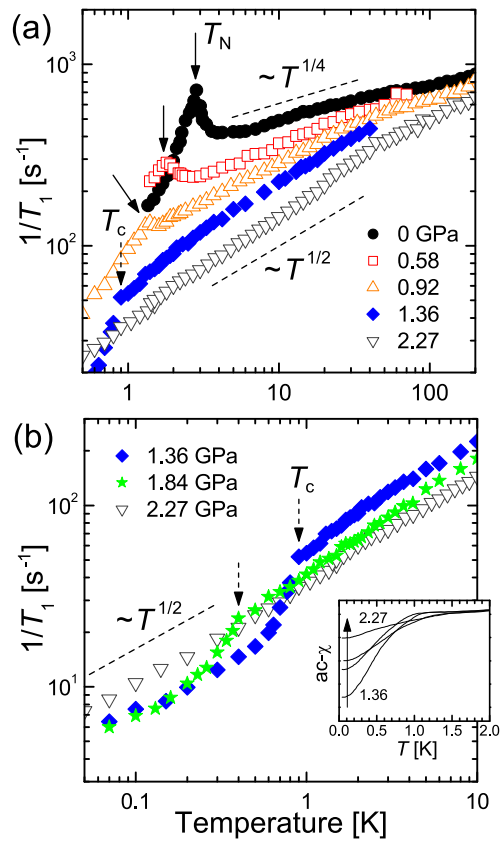}
\caption[]{\footnotesize (Color online) $T$ dependences of $1/T_1$ at (a) high $T$ and (b) at low $T$ for $P=0-2.27$ GPa in Ce$_2$RhIn$_8$. Solid and dashed arrows point to $T_N$ and $T_c$, respectively. The inset shows the $T$ dependence of ac-susceptibility at $P$ = 1.36, 1.62, 1.84, and 2.27 GPa in the order indicated by the direction of the arrow.}
\label{T1}
\end{figure}

\begin{figure}[htbp]
\centering
\includegraphics[width=6.5cm]{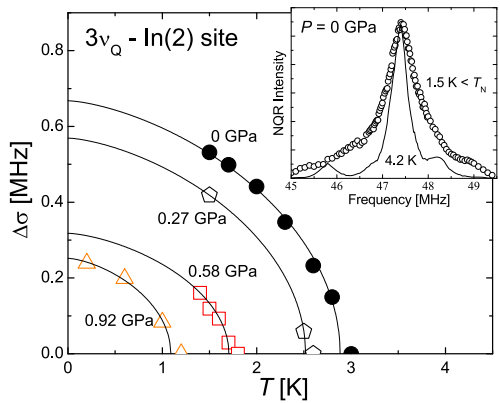}
\caption[]{\footnotesize (Color online) The $T$ dependence of $\Delta\sigma(T)$ at In(2) in Ce$_2$RhIn$_8$ for several pressures. The solid lines represent the relation $\Delta \sigma(T)\propto [ 1-(T/T_N)^{3/2} ]^{1/2}$. The inset shows the NQR spectra above and below $T_N$ at $P$ = 0 GPa.}
\label{NQR}
\end{figure}

\section{I\hspace{-.1em}I\hspace{-.1em}I. RESULTS AND DISCUSSION}

Fig.~\ref{T1}a shows the $T$ dependence of $1/T_1$ at high $T$ and $P=0-2.27$ GPa in Ce$_2$RhIn$_8$. A distinct peak in $1/T_1$ is associated with the onset of AFM order at $T_N=2.85$ K and $P$ = 0 GPa. It was reported from the resistivity measurements that the secondary anomaly ($T_{LN}$) well below $T_N$ was observed in the vicinity of ambient $P$ \cite{Nicklas}. However, it was not observed from the present NQR measurements either, as reported in the previous NQR paper by Fukazawa $et~al$. \cite{Fukazawa}. Note that in the PM state, $1/T_1$ increases up to 200 K at $P=0$, suggesting that Ce-derived magnetic fluctuations occur in an itinerant regime; this is consistent with the NQR measurement results \cite{Fukazawa} and the angle-resolved photoemission spectroscopy results \cite{Raj}. The behavior $1/T_1\propto T^{1/4}$ is consistent with a quasi-2D AFM spin-fluctuations (SFs) model that predicts the relation $1/T_1T\propto\chi_Q(T)^{3/4}$ near an AFM QCP \cite{Lacroix}. Here, the term quasi-2D AFM SFs implies that the magnetic correlation length in the tetragonal plane develops at a faster rate than that along the c-axis and that the staggered susceptibility $\chi_Q(T)$ with the AFM wave vector $Q=(1/2, 1/2, 0)$ is anticipated to obey the Curie-Weiss law as $\chi_Q(T)\propto 1/(T+\theta)$. In this context, it is predicted that the quasi-2D AFM SFs will obey $1/T_1\propto T\times\chi_Q(T)^{3/4}\propto T^{1/4}$ in the vicinity of the AFM QCP, where $\theta \sim 0$. As $P$ increases, the $T_N$ determined from a peak in $1/T_1$ decreases to $T_N$ = 1.2 K at $P$ = 0.92 GPa. At $P$ = 1.36 GPa, a marked decrease in $1/T_1$ below 0.9 K without an accompanying peak was observed, which was unexpected. As mentioned later, this is because SC sets in below $T_c=0.9$ K.

Next, we deal with the possible existence of the AFM-QCP in Ce$_2$RhIn$_8$. The inset in Fig.~\ref{NQR} shows the 3$\nu_Q$-NQR spectra corresponding to In(2) above and below $T_N$ at $P=0$. The main peak inherent to In(2) in Ce$_2$RhIn$_8$ is accompanied by two satellite peaks at $\sim$ 45.8 and $\sim$ 48.2 MHz, which are thought to be due to stacking faults in the Ce$_2$RhIn$_8$ that consists of alternating layers of CeRhIn$_5$ and CeIn$_3$ since the spectral intensities of these peaks are almost negligible. In fact, the X-ray diffraction measurements revealed a diffuse scattering suggesting stacking faults along the c-axis of Ce$_2$RhIn$_8$ \cite{Koeda}. The full width at the half maximum $\sigma(T)$ of the 3$\nu_Q$-NQR spectrum increases due to $H_{\rm int}$ induced by the AFM moments that develop below $T_N$. Figure~\ref{NQR} shows the $T$ dependence of $\Delta\sigma(T)$ at In(2) in Ce$_2$RhIn$_8$ for several pressures. Here, $\Delta \sigma(T)=\sigma(T)-\sigma(T_N)$, which is approximately proportional to the magnitude of the AFM ordered moment. At $P=0$, $\Delta \sigma(T)$ is well fitted by the relation $\Delta \sigma(T)\propto [ 1-(T/T_N)^{3/2} ]^{1/2}$, which is expected in a weak itinerant AFM \cite{Nakayama,Hasegawa}, as indicated by the solid line in Fig.~\ref{NQR}. Using this relation for $\Delta \sigma(T)$ under $P$, we tentatively estimate $\Delta \sigma(T=0)$, as shown in Fig.~\ref{summary}a. Note that as $P$ increases, $\Delta \sigma(T=0)$ decreases linearly and a rough extrapolation to $\Delta \sigma$ = 0 yields $P_{\rm QCP}\sim$ 1.36 GPa.  Furthermore, note that as $P$ increases, the behavior $1/T_1\propto T^{1/4}$ at $P$ = 0 evolves into $1/T_1\propto T^{1/2}$ around $P_{\rm QCP}$, as shown in Fig.~\ref{T1}a. The latter relation is consistent with the 3D-AFM SFs model that predicts the relation $1/T_1T \propto \chi_Q(T)^{1/2}$ near the 3D-AFM QCP \cite{Moriya}. When assuming a simple power-law dependence for $1/T_1$, e.g., $1/T_1=AT^n$ with parameters $A$ and $n$, the systematic $T$ variations of $1/T_1$ are fitted in the $T$-range from $T$ well above $T_N$(or $T_c$) to 30 K to obtain the $P$ dependence of $n$, as shown in Fig.~\ref{summary}a. Note that $n$ progressively increases up to $n=0.5$ at $P_{\rm QCP}=1.36$ GPa and remains almost constant as $P$ increases further, indicating that the crossover from the quasi-2D to 3D character of AFM SFs occurs between $P$ = 0 and 1.36 GPa.

\begin{figure}[htbp]
\centering
\includegraphics[width=6.5cm]{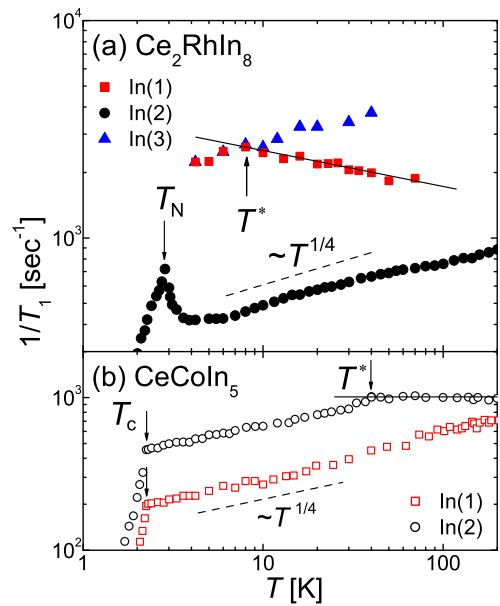}
\caption[]{\footnotesize (Color online) (a) The $T$ dependence of $1/T_1$ for the In(1), In(2), and In(3) sites at ambient $P$ in Ce$_2$RhIn$_8$. (b) The $T$ dependence of $1/T_1$ for the In(1) \cite{Yashima2} and In(2) sites at ambient $P$ in CeCoIn$_5$. The solid lines are eye-guides. $T^*$ is the tempearature at which the anomaly in the $T$ dependence of 1/$T_1$ appears.}
\label{HyperFine}
\end{figure}

Previous papers reported that $1/T_1$ at In(1) differs from that at In(2) \cite{Fukazawa}. We have confirmed that $1/T_1$ at In(3) resembles the corresponding result for In(2), but above $T^* \sim$ 8 K, $1/T_1$ at In(1) deviates from the $T^{1/4}$ behavior, as shown in Fig.~\ref{HyperFine}a. Note that the In-site dependence of $1/T_1$ was also observed in CeCoIn$_5$, as shown in Fig.~\ref{HyperFine}b. This is understood in terms of the $T$ dependence of the hyperfine-coupling constants at In sites under a crystal electric field (CEF) effect. As a matter of fact, the NMR study reported by Curro $et~al$. revealed that the energy splitting between the first excited CEF level and the ground state ($\Delta_{CEF}$) is estimated at 34 K and hence the hyperfine couplings at In(2) significantly changes around 50 K close to $T^* \sim$ 40 K \cite{Curro}. Likewise, since $\Delta_{CEF}$ in Ce$_2$RhIn$_8$ is estimated at 14 K that was deduced from the magnetic susceptibility and magnetization measurements \cite{Ueda}, the hyperfine couplings at In(1) in this compound may change around a temperature close to $T^* \sim$ 8 K.

\begin{figure}[htbp]
\centering
\includegraphics[width=8cm]{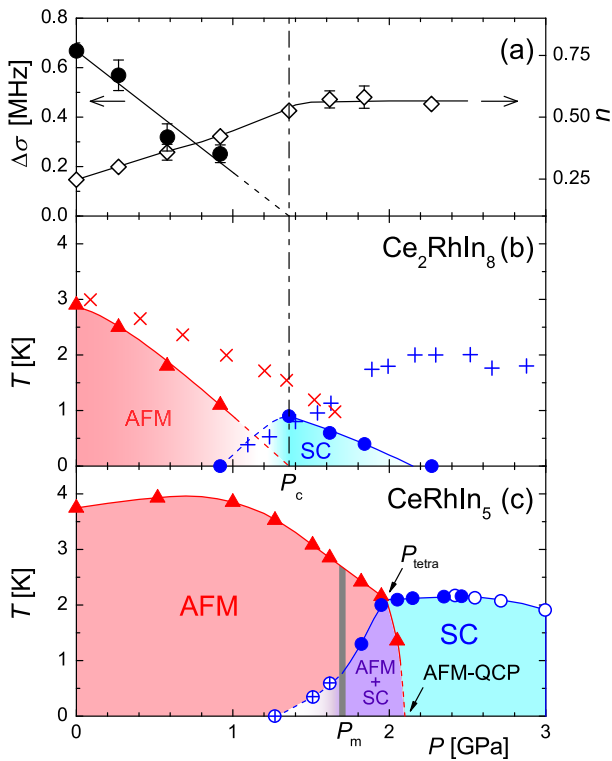}
\caption[]{\footnotesize  (Color online) (a) The $P$ dependence of $\Delta\sigma(T=0)$ at In(2) and $n$ where $1/T_1 \propto T^n$ in the PM state at temperatures well above $T_N$(or $T_c$). (b) The $P-T$ phase diagram of Ce$_2$RhIn$_8$. The data denoted by cross and plus marks indicate the $P$ dependences of $T_N$ and $T_c$, as determined from heat-capacity \cite{Lengyel} and resistivity\cite{Nicklas} measurements, respectively. (c) The $P-T$ phase diagram of CeRhIn$_5$ \cite{Yashima1,Yashima3}. The commensurate AFM is completely realized above $P_m$.}
\label{summary}
\end{figure}

In order to demonstrate the onset of SC in Ce$_2$RhIn$_8$, in Fig.~\ref{T1}b, we present the $T$ dependences of $1/T_1$ at low $T$ and $P$ = 1.36, 1.84 and 2.27 GPa, where the AFM order collapses. Although the onset of SC is proved by the appearance of SC diamagnetism, as indicated in the inset in Fig.~\ref{T1}b, this diamagnetism cannot be used to identify a transition temperature $T_c$ for bulk SC inherent to Ce$_2$RhIn$_8$ under $P$. In fact, the SC diamagnetism for $P >$ 1.84 GPa starts to appear from a relatively high $T$ onwards. This may be associated with the diamagnetism arising from the CeRhIn$_5$ contained in the sample as an impurity phase. This CeRhIn$_5$ contamination leads to inconsistencies among $P-T$ phase diagrams, depending on the experimental methods \cite{Nicklas,Ohara,Ueda,Lengyel}. On the other hand, a marked reduction in the $T$ dependence of $1/T_1$, which is shown in Fig.~\ref{T1}b, provides microscopic evidence for the development of SC in the sample at $T_c=0.9$ and 0.4 K and $P$ = 1.36 and 1.84 GPa, respectively. In contrast, the $1/T_1$ value at 2.27 GPa does not yield such evidence, though the diamagnetism starts to appear below $\sim$ 1.5 K. Thus, material-selective NQR-$T_1$ measurements allow us to identify the onset of the SC inherent to Ce$_2$RhIn$_8$ under $P$. It is remarkable that significantly large diamagnetism and SC with $T_{c}^{max}=0.9$ K are observed at $P_{\rm QCP}=1.36$ GPa. These results suggest the intimate relationship between the unconventional SC and the AFM QCP in Ce$_2$RhIn$_8$. Furthermore, it should be noted that SC sets in as a result of the evolution from the quasi-2D to 3D character of AFM SFs. This is in contrast to the fact that the SC dome in CeCoIn$_5$ and CeRhIn$_5$ with $T_c^{max} >$ 2 K is realized around the quasi-2D AFM QCP but is separated from the phase boundary between the AFM and PM phases. These results demonstrate the intimate relationship between the dimensionality of AFM SFs and the onset of unconventional SC; the 2D character of AFM SFs is favorable to the increase of the $T_c$ in HF SC compounds as well as in high-$T_c$ copper oxides \cite{Scott}.

As an indication that the symmetry of the SC gap function in Ce$_2$RhIn$_8$ must be considered, we note that $1/T_1$ at $P$ = 1.36 GPa decreases without the appearance of a coherence peak just below $T_c$ and exhibits a large kink well below $T_c$, associated with the existence of the large residual density of states. These results suggest a dirty $d$-wave SC with line-nodes gap, identical to the case of high-$T_c$ superconductors \cite{Ishida}. This may be because difficulties in preparing the crystals containing alternating layers of CeRhIn$_5$ and CeIn$_3$ lead to impurities and/or crystal imperfections like stacking faults in Ce$_2$RhIn$_8$. It is well known that the existence of the residual density of states due to the impurity effect results in $T$-linear behavior well below $T_c$. Unexpectedly, however, the observed behavior $1/T_1\propto T^{1/2}$ well below $T_c$ cannot be simply explained by the impurity effect for unconventional SC; this indicates the persistence of low-lying excitations in the SC state due to the proximity to the AFM QCP. The enhancement of $1/T_1$ even at temperatures lower than $T_c$ is also observed in the uniformly coexisting state of SC and AFM around the AFM-QCP in CeCu$_2$Si$_2$ \cite{YKawasaki1}, CeRhIn$_5$ \cite{Yashima3}, CeCo(In$_{1-x}$Cd$_x$)$_5$ \cite{Urbano}, and CeNiGe$_3$ \cite{Harada}. However, note that in Ce$_2$RhIn$_8$, the behavior of $1/T_1\propto T^{1/2}$ is observed even in the SC state where the AFM order collapses. In this context, the $P-T$ phase diagram for Ce$_2$RhIn$_8$ is the only one that reveals the following unconventional SC characteristic: 3D-AFM SFs survive in the SC state that occurs in the relatively narrow $P$ range 1.36 -- 1.84 GPa.

\section{I\hspace{-.1em}V. CONCLUSION}

In conclusion, we have established the $P-T$ phase diagram for Ce$_2$RhIn$_8$ from microscopic In-NQR measurements. The AFM order disappears at $P_{\rm QCP}\sim$ 1.36 GPa, where 3D-AFM SFs are dominant. It was demonstrated that the SC order occurs in the narrow $P$ range of 1.36 -- 1.84 GPa and exhibits $T_c^{max}$ = 0.9 K around $P_{\rm QCP}\sim$ 1.36 GPa. We state that this phase diagram differs from the previously reported ones \cite{Bao} because the latter were affected by contamination by impurity phases such as CeRhIn$_5$. The unconventional SC in Ce$_2$RhIn$_8$ occurs under the development of 3D AFM SFs rather than the quasi-2D AFM SFs, as in the case of CeCoIn$_5$\cite{Sidorov,YKawasaki2,Yashima2} and CeRhIn$_5$\cite{Muramatsu}. Noting that the $T_c^{max}$ (= 0.9 K) for Ce$_2$RhIn$_8$ is significantly lower than the $T_c(>$ 2 K) for CeCoIn$_5$ and CeRhIn$_5$, it is suggested that the 2D character of AFM SFs plays a vital role in increasing the $T_c$ in strongly correlated electron systems.

\section{ACKNOWLEDGMENTS}

This work was supported by Grants-in-Aid for Specially Promoted Research (Grant No. 20001004) and for Young Scientists (B) (Grant No. 20740195) from the Ministry of Education, Culture, Sports, Science and Technology (MEXT) of Japan. It was partially supported by the Global COE Program (Core Research and Engineering of Advanced Materials-Interdisciplinary Education Center for Materials Science) from MEXT.

\end{document}